\documentclass[aps,rmp,twocolumn,footinbib,notitlepage]{revtex4-1}
\usepackage[a4paper,top=1.5cm,bottom=1.5cm,left=1.75cm,right=1.75cm]{geometry}
\usepackage{enumerate}
\usepackage{amsmath}
\usepackage{amssymb}

\usepackage{overpic}
\usepackage{tikz}
\usepackage{pgfplots}
\usepackage{bm}
\usepackage{mathrsfs}

\usepackage{mathabx} 
\usepackage{graphicx}
\DeclareGraphicsExtensions{.eps, .pdf, .jpg, .png}
\graphicspath{{./figs/}}
\usepackage{hyperref}
\def\ns{\negthickspace}
\newcommand{\kb}{k_{\text{B}}} 
\def\vq{{\bf q}}
\def\vz{{\bf z}}
\def\vr{{\bf r}}
\def\vn{{\bf n}}
\def\vA{{\bf A}}
\def\vB{{\bf B}}
\def\vE{{\bf E}}
\def\vJ{{\bf J}}
\def\cC{{\mathcal C}}
\renewcommand{\Im}{\mbox{Im\,}}
\renewcommand{\Re}{\mbox{Re\,}}
\newcommand{\grad}{\bm\nabla}
\renewcommand{\onlinecite}[1]{\hspace{-1 ex} \nocite{#1}\citenum{#1}}
\setcitestyle{numbers,super}
\begin{document}
\title{Electromagnetic Response of Disordered Superconducting Cavities}
\author{Mehdi Zarea}
\email{zarea.mehdi@gmail.com}
\author{Hikaru Ueki}
\email{uekih@lsu.edu}
\author{J. A. Sauls}
\email{sauls@lsu.edu}
\affiliation{Hearne Institute of Theoretical Physics,
             Department of Physics \& Astronomy,
             Louisiana State University,\\
             Baton Rouge, LA 70803 USA}
\date{\today}
\begin{abstract}
We present results for the resonant frequency shift and quality factor of disordered Nb SRF cavities driven out of equilibrium by the resonant microwave field. 
The theory is based on the nonequilibrium theory of superconductivity for the current response to the electromagnetic field at the vacuum-metal interface. 
We are able to accurately predict the observed frequency shifts with a precision of order several Hz over the full temperature range $0 < T \le T_c$, including the negative frequency shift anomalies that are observed very near $T_c$. The origin of these anomalies is shown to be the competition between the normal metal skin depth and the London penetration depth which diverges as 
$T\rightarrow T_c^-$. An analytic approximation to the full current response, valid for $|T-T_c|\ll T_c$, accounts for the negative frequency shift near $T_c$. The non-monotonic dependence of the quality factor on the quasiparticle scattering rate is related to the pair-breaking effect of disorder on the superfluid fraction, and thus the London penetration depth. 
\end{abstract}
\maketitle
\section{Introduction}
\label{sec-Introduction}

Superconducting radio-frequency (SRF) cavities made of Niobium are a key technology for high energy particle accelerators. Materials processes such as Nitrogen infusion combined with heat treatments have led to significant improvements in cavity quality factor now of order $Q\approx 2\times 10^{11}$ as well as increased accelerating gradients now approaching $50\,\mbox{MV/m}$.~\cite{gra13,gra17}
High Q SRF cavities provide a novel platform as sensors for rare events, e.g. as detectors for photon-photon scattering at microwave frequencies mediated by virtual electron-positron pairs, or by pseudo-scalar axions.~\cite{bog19,gao21}
High Q SRF cavities have also been proposed as exceptional resonators for quantum memory with photon lifetimes exceeding $T_1\approx 2\,\mbox{sec}$, and as quantum processors.~\cite{rom20}
The performance of these cavities, for both accelerator applications or as quantum devices, is sensitive to the surface screening currents and the impact of surface and bulk disorder on the current response.~\cite{gur17,nga19,uek22,sau22}
The sensitivity of the screening current to disorder is highlighted by measurements of the frequency shift of N-doped Nb SRF cavities as a function of temperature, for four cavities with frequencies $f=0.65, 1.3, 2.6, 3.9~GHz$.~\cite{baf21}
All four cavities show a negative frequency shift confined to a narrow range of temperatures near the transition temperature $T_c$, followed by a rapid rise to a positive frequency shift that satures at low temperature.
The non-monotonic temperature dependence (``anomaly'') of the negative shift of the resonant frequency of the cavity just below $T_c$, i.e. $\delta f=f_s(T)-f_n(T_c)$, for $|T-T_c|\ll T_c$ is sensitive to surface treatment and to disorder in the region of the screening currents.~\cite{baf21}
A similar anomaly in the frequency shift just below $T_c$ was reported much earlier for Nb coupled to a tunnel-diode oscillator operating at $f\approx 10\,\mbox{MHz}$.~\cite{var75}
In addition, analysis of the surface impedance data of Niobium embedded in a cylindrical copper cavity~\cite{kle94a} implies a negative shift in the resonance frequency shift at $f=60\,\mbox{GHz}$.~\cite{uek22}
Thus, the resonance frequency of oscillators made of, or coupled to, superconducting Nb exhibit a negative shift anomaly for resonant frequencies spanning nearly four orders of magnitude, albeit in all cases with $h f \ll 2\Delta(0)\approx 3.6\,\kb T_c$. 
Nevertheless, the temperature of the maximum negative shift, as well as the magnitude of the negative frequency shift depend on both frequency, $\omega$, and disorder, the latter parametrized by the quasiparticle-impurity mean scattering time, $\tau$.

In Ref.~\onlinecite{uek22} we developed the theory for the complex surface impedance $Z_s = R_s + i X_s$ and its connection to the resonant frequency and quality factor of SRF cavities based on Slater's approach to solving Maxwell's equations for enclosed electromagnetic cavities~\cite{sla46} in terms of surface and volume responses of the cavity wall and dielectic medium within the cavity, which in our case is N-doped superconducting Nb for the cavity wall and vacuum for the dielectric medium.

In this report we derive an equation for the complex eigenfrequency for the lowest TM mode of a cylindrical RF cavity including penetration of the EM field and its confinement by the normal and superconducting currents in the vicinity of the vacuum-metal interface.   
The current response is obtained from the Keldysh formulation of the quasiclassical theory of superconductivity.~\cite{rai94b}
The resulting eigenvalue equation is straight-forward to solve numerically, as well as analytically in certain limits. From the complex eigenvalue equation we calculate both the quality factor, $Q$, and frequency shift, $\delta f$, of the cavity as a function of cavity geometry, fundamental frequency, $\omega$, temperature, $T$, and material properties of the superconductor such as $T_c$, scattering rate from the disorder potential, $\hbar/\tau$, etc.
Here we report results based on in the low-field linear response limit for the screening current at microwave frequencies. The main results are: (1) the eigenvalue equation for the fundamental mode of an SRF cavity based on our theory of the superconducting state of disordered Niobium,~\cite{zar22} (2) theoretical results for $\delta f$ as a function of temperature for disordered Nb and the quantitative comparison with experimental results reported for an N-doped Nb SRF cavity, (3) an approximate analytical result for the non-monotonic, negative frequency shift ``anomaly'' that is characteristic of sufficiently disordered SRF cavities for temperatures very close to $T_c$, and (4) non-monotonic dependence of $Q$ as a function of the mean quasiparticle-impurity scattering rate, $1/\tau$, with a maximum near $\tau\sim\hbar/2\pi\kb T_c$, i.e. the pair formation timescale.

\section{Current Response of an SRF Cavity}\label{sec-EM-response}

SRF cavities are open quantum systems in which photons in a single mode are coupled to the environment of a superconducting metal - normal electrons, Cooper pairs, phonons, as well as embedded impurities and two-level defects - which confines the photons within the cavity for relatively long timescales, $T_1\simeq 1\,\mbox{sec}$, for Nb SRF cavities with $f\simeq \mbox{GHz}$ and $Q\simeq 10^{11}$.
The penetration of the EM field into the superconductor also shifts the resonant frequency of the cavity by $\delta f\sim 0.1 - 50\,\mbox{kHz}$.
It is possible to predict the magnitude and the variations of the resonant frequency and quality factor of Nb-based SRF cavities with remarkable precision.

We first solve Maxwell's equations for the EM field with the constitutive equation for the screening current response to a transverse EM field within the superconductor, 
\begin{equation}
\label{eq-Current_response}
\vJ(\vr,t) = -\frac{c}{4\pi}\int_{-\infty}^{t}\,K(t-t')\,\vA(\vr,t')
\,,
\end{equation}
where $K^{R}(t-t')\equiv\Theta(t-t')\,K(t-t')$ is the retarded response function. For the analysis in this report we assume there is sufficient disorder that the current response to $\vA(\vr,t)$ can be evaluated in the local limit.
Thus, the Amp\`ere-Maxwell equation for the EM field in the superconductor becomes,
\begin{equation}
\label{eq-Maxwell-London_Equation}
\left(
\frac{1}{c^2}\partial^2_t - \nabla^2
\right)\vA(\vr,t) 
= 
-
\int_{-\infty}^{t}\,K(t-t')\,\vA(\vr,t')
\,,
\end{equation}
where we work in the gauge $\grad\cdot\vA=0$ for a purely transverse EM field.
Inside the cavity $\vA(\vr,t)$ satisfies the wave equation in vacuum. Thus, to obtain the full solution for the EM field in the cavity and the surface region of the cavity walls we must solve the free field wave equation and Eq.~\eqref{eq-Maxwell-London_Equation}, with the boundary condition for the continuity of $\vA$ and the normal derivative at the vacuum-superconductor interface.  
For a resonant mode of frequency $\omega$ Eq.~\eqref{eq-Maxwell-London_Equation} 
reduces to,
\begin{equation}
\label{eq-Maxwell-London_Equation-omega}
\left(
\nabla^2
+
\frac{\omega^2}{c^2}
-
K(\omega)
\right)\vA(\vr,\omega) = 0
\,,
\end{equation}
where $K(\omega)$ is the Fourier transform of the retarded current response function appearing in Eq.~\eqref{eq-Current_response}.

\medskip
\subsection{Keldysh Response Function}

The linear response function $K(\vq,\omega)$ for superconductors subject to excitation by an electromagnetic field is calculated using Keldysh's formulation of nonequilbrium response in the quasiclassical approximation.~\cite{rai94b} The current response depends on the frequency and wavevector, $\omega$ and $\vq$, of the electromagnetic field within the metal, includes the condensate response (supercurrent) and the dissipative response of unbound quasiparticles scattered by the random potential with scattering rate, $1/\tau$.
The other important internal timescale is the pair formation time, $\tau_0\equiv\hbar/2\pi\kb T_c$, and the corresponding ballistic pair correlation length is $\xi_0=v_f\tau_0=\hbar v_f/2\pi\kb T_c$. For pure Nb the pair correlation length and London penetration depth are comparable, $\xi_0\approx\lambda_{\text{L$_0$}}\simeq 33\,\mbox{nm}$.
However, disorder leads to a finite mean free path for quasiparticles, $\ell=v_f\tau$, converting ballistic motion to diffusive motion, and thus a reduction in the pair correlation length. Disorder is also pair breaking, particularly in the presence of screening currents, resulting in a reduction of the superfluid density, $n_s$. Thus, the London penetration depth, $\lambda_{\text{L}}\propto 1/\sqrt{n_s}$, increases due to scattering by the random potential.

For the analysis reported here we consider the long-wavelength (``local'') limit, $q\lesssim 1/\lambda_{\text{L}}\ll\,1/\xi_0$, for the current response, which is achieved in Nb when the mean scattering time is comparable to the pair formation time, $\tau\sim\tau_0$.
This level of disorder leads to weak suppression of $T_c$, a reduction in the superconducting coherence length, $\xi$, and an increase in the London penetration depth such that $\lambda\gg\xi$.
In this limit the screening current is determined by the local value of the electromagnetic field, and we can set $q=0$ in the current response function. The result is given by~\cite{rai94b}
\begin{widetext}
\hspace*{-7mm}
\begin{eqnarray}
\label{eq-K}
K(\omega;\tau,T)
\ns=\ns
\frac{\pi\sigma_{\text{D}}}{ic^2\tau}\
\int_{-\infty}^{+\infty}\ns\ns d\epsilon\ns\ns
&\Bigg\{&
\tanh\left(\frac{\epsilon-\omega/2}{2T}\right)
\frac{1}{D^R_+ + D^R_- + 1/\tau}
\left(\frac{\epsilon^2+\Delta^2-\omega^2/4}{D_+^R D_-^R}+1\right) 
\nonumber\\
&-&
\tanh\left(\frac{\epsilon+\omega/2}{2T}\right)
\frac{1}{D^A_+ + D^A_- + 1/\tau}
\left(\frac{\epsilon^2+\Delta^2-\omega^2/4}{D_+^A D_-^A}+1\right) 
\,
\\ 
&+&
\left[
\tanh\left(\frac{\epsilon+\omega/2}{2T}\right)
-
\tanh\left(\frac{\epsilon-\omega/2}{2T}\right)
\right]
\frac{1}{D_+^R + D_-^A + 1/\tau}
\left(\frac{\epsilon^2+\Delta^2-\omega^2/4}{D_+^R D_-^A}+1\right) 
\Bigg\}
\,,
\nonumber
\end{eqnarray}
\end{widetext}
where $\sigma_{\text{D}}=\frac{2}{3}N_f e^2 v_f^2 \tau = n e^2 \tau/m^*$ is the Drude result for the d.c. conductivity for charge $e$ quasiparticles, with Fermi velocity, $v_f$, Fermi momentum $p_f$, density of states, $N_f={\tiny \frac{3}{2}}n/p_fv_f$, where $n$ is the electron density and $m^*=p_f/v_f$ is the quasiparticle effective mass.
The denominators in Eq.~\eqref{eq-K} are defined by
\begin{equation}
\label{eq-DRA}
D^{R/A}_{\pm}
\equiv \sqrt{\Delta^2 -(\epsilon \pm \omega/2 \pm i\delta)^2}
\,,
\end{equation}
and are retarded (advanced) functions defined by the analytic continuation to the real axis indicated by $+i\,\delta$ ($-i\,\delta$) and $\delta\rightarrow 0^+$.
The response function $K(\omega)$ is directly related to the bulk microwave conductivity,
\begin{equation}\label{eq-conductivity}
\sigma=\sigma_1+i\sigma_2
=i\frac{c^2}{4\pi\omega}\,K(\omega)
\,,
\end{equation}
where $\sigma_1=\Re\sigma(\omega)$ and $\sigma_2(\omega)=\Im\sigma(\omega)$ 
are the real and imaginary parts of the a.c. conductivity.
In the normal state, $\Delta\rightarrow 0$, the conductivity, $\sigma(\omega)$, defined by Eqs.~\eqref{eq-K} and~\eqref{eq-conductivity} reduces to
\begin{equation}
\sigma_n(\omega) = 
\frac{\sigma_{\text{D}}}
{1-i\omega\tau}
\,.
\end{equation}

Superconductivity leads to significant changes in the the cavity resonance frequency, as well as the quality factor, that are sensitive to disorder and temperature, which we calculate to predict and analyze experimental data using Eqs.~\eqref{eq-Maxwell-London_Equation-omega},~\eqref{eq-K},~\eqref{eq-DRA} and boundary conditions for the EM field at the vacuum-metal interface. 

\begin{table}
\label{table-material_parameters}
\begin{center}
\begin{tabular}{|c|c|c|c|c|c|}
\hline
$T_c\,[\mbox{K}]$ 
& 
$v_f\,[10^8\,\mbox{cm/s}]$
& 
$\tau_0\,[\mbox{ps}]$ 
& 
$\xi_0\,[\mbox{nm}]$ 
& 
$\lambda_{\text{L$_0$}}\,[\mbox{nm}]$ 
& $\Delta_0\,[\mbox{meV}]$ 
\\
\hline
9.33 
& 
0.257 
& 
0.131
& 
33.0 
& 
33.0 
& 
1.55 
\\
\hline
\end{tabular}
\caption{Material parameters for pure Niobium. 
Note that 
$\tau_0=\hbar/2\pi\kb T_{c_0}$, 
$\xi_0=v_f\tau_0$,
$\lambda_{\text{L$_0$}} = c/\omega_p$
and 
$\Delta_0$ is the strong-coupling gap for pure Nb.~\cite{zar22}
}
\end{center}
\end{table}

\subsection{Cylindrical cavities}

SRF cavities for accelerator applications adopt the Tesla geometry, i.e. axially symmetric with an oval shape designed in part to eliminate sharp corners which are sources of field emission.~\cite{padamsee08}
Here we consider cylindrical cavities of radius $R$ and length $L$. This model is chosen in order to simply our theoretical analysis, in particular in implementing the boundary conditions on the EM field for the lowest TM mode at the vacuum-metal interface.
Theoretical results depend on the response function, the local boundary conditions on the field at the vacuum-metal interface, and an overall geometric factor, $G$.
The latter geometric factor depends on the field distribution within the geometry of the cavity and is independent of the temperature and material properties of the cavity walls. Thus, we correct our result obtained using the TM$_{010}$ mode of the cylindrical cavity by replacing the value of $G$ for the cylindrical cavity with the geometric factor computed for the lowest TM mode of the the Tesla cavity.

For the cylindrical cavity the lowest frequency TM resonance is the TM$_{010}$ mode with the vector potential along the cylinder axis, $\vA(\vr,\omega)=A_0\,J_0(\omega\rho/c)\,\hat{\vz}$ in the gauge $\grad\cdot\vA=0$. The Fourier component of the electric field is axial, $\vE(\vr,\omega)=(i\omega/c)\,A_0\,J_0(\omega\rho/c)\,\hat{\vz}$, and the magnetic field circulates azimuthally, $\vB(\vr,\omega)=A_0\,(\omega/c)\,J_1(\omega\rho/c)\,\hat{\pmb\varphi}$.
The perfect conductor boundary condition, $\hat\vn\times\vE\vert_{S}=0$, where $\hat\vn$ is a local unit vector normal to any point on the surface $S$, determines the field distribution in the interior to the cavity and the eigenfrequency of the cavity based on the cavity geometry. 
In particular, we have $J_0(\omega R/c)=0$, and thus, a resonant frequency of $\omega=x_{01} c/R$ where $x_{01}\simeq 2.405$ is the first zero of $J_0(x)$. The electric field is maximum along the axis of the cavity and vanishes on the cylinder wall, while the magnetic field vanishes along the axis of the cavity and is a maximum on the cylinder wall, $B_{\text{max}}=\cC\,A_0 R$, where $\cC\equiv x_{01}\,J_1(x_{01})\simeq 1.248$.
The geometric factor for a cylindrical cavity of length $L$ and radius $R$ reduces to,~\cite{padamsee08} 
\begin{equation}
\label{eq-G_cyl}
G=Z_0\,(\omega/c)\,
\frac{
\int_{V}d^3r\,\vert\vB(\vr,\omega)\vert^2
}
{\int_{S}dS\,\vert\vB(\vr_S,\omega)\vert^2
}
=Z_0\,\frac{x_{01}}{2}\,\frac{L}{L+R}
\,,
\end{equation}
where $Z_0=4\pi/c\approx 377\,\Omega$ is the vacuum impedance. 

However, for a vacuum-superconductor interface the vector potential, and thus a tangential electric field as well as tangential magnetic field, penetrates into the superconductor, but is confined to the interface within a distance of order the London penetration depth, of order $\lambda_{\text{L}}\simeq 50\,\mbox{nm}$ for moderately disordered Niobium.
The penetration of the field over distance scales of order $\lambda_{\text{L}}\ll R$ leads to a small shift in the resonance frequency, $\delta\omega$, which depends on the material properties of the superconductor via the current response function $K(\omega)$. Penetration of the electric field into the superconductor also leads to dissipation of the microwave field by quasiparticles scattering from the interface and from the random distribution of impurities, thus limiting the quality factor, $Q$.

Both $\delta\omega$ and $Q$ are determined by an eigenvalue equation obtained from the boundary conditions on the vector potential and its derivative at the vacuum-superconductor interface. 
In particular, for the TM$_{010}$ mode on the vacuum side, $A(\rho)=A_0\,J_0(\frac{\omega}{c}\rho)$. However, on the superconducting side the confinement of the field to distances of order $\lambda_{\text{L}}\ll R$ allows us to neglect the curvature of the interface. Thus, for $\rho\ge R$ we obtain $A(\rho)=A'_0\,\exp\{-\Lambda(\omega)(\rho-R)\}$, where $\Lambda(\omega)\equiv\sqrt{K(\omega)-(\omega/c)^2}$.
The ratio of the two boundary conditions reduces to
\begin{equation}
\label{eq-eigenvalue_equation}
\hspace*{-3mm}
\Lambda(\varpi)
\ns=\ns
\sqrt{K(\varpi) - (\varpi/c)^2}
\ns=\ns
\left(\frac{\varpi}{c}\right)
\frac{J_1(\varpi R/c)}{J_0(\varpi R/c)}
\,,
\end{equation}
where $J_1(x)=-dJ_0(x)/dx$.
Note that the response function is in general a complex function of $\omega$, which is analytic in upper half of the complex frequency plane, thus $\omega\rightarrow\varpi$, with $\Im\varpi>0$.
Equation~\eqref{eq-eigenvalue_equation} is the key equation for the complex eigen-frequency, $\varpi$, that determines the resonance frequency, the penetration depth and the quality factor, all of which become functions of temperature, disorder and frequency.
The results reported here support and agree with our analysis of disordered SRF cavities based on Slater's method.~\cite{uek22}

\subsection{Field Penetration and Frequency Shifts}

Before presenting numerical results for the frequency shift and quality factor of disordered SRF cavities, we discuss the physics underlying the dependence of the cavity resonance frequency on disorder and temperature in both the normal and superconducting states.

In the limit $\omega\tau\ll 1$, $\sigma_{n}\simeq\sigma_{\text{D}}$ determines the dissipation of microwave power and the penetration depth of the EM field in the normal metal. In particular, on the metallic side of the vacuum-metal interface ($x>0$),
\begin{equation}
A_n(x,\omega) = A_0\,\Re e^{iq(\omega)x}
\,,
\end{equation}
where $q(\omega)=\sqrt{i\,4\pi\sigma_{\text{D}}\omega}/c=(1+i)/\delta(\omega)$ with
\begin{equation}\label{eq-delta}
\delta(\omega)\equiv c/\sqrt{2\pi\sigma_{\text{D}}\omega}
=
\frac{\lambda_{\text{L$_0$}}}{\sqrt{\omega\tau/2}}
\,.
\end{equation}
Note that $\lambda_{\text{L$_0$}}=c/\omega_{p}$ is the clean limit result for the London penetration depth at zero temperature, defined here in terms of the plasma frequency, $\omega_p\equiv\sqrt{{\tiny\frac{8\pi}{3}} N_f e^2 v_f^2}$.
The penetration of the EM field in the normal metal is then given by
\begin{equation}\label{eq-lambda_n}
\lambda_n = \frac{1}{A_0}\int_{0}^{\infty}dx\,A_n(x,\omega) = \Re\left(\frac{i}{q(\omega)}\right) = \frac{\delta(\omega)}{2}
\,.
\end{equation}
In the superconducting state, $A_s(x,\omega)=A_0\,e^{-x/\lambda_{\text{L}}}$, and thus the penetration depth of the EM field is given by the London length,
\begin{equation}
\lambda_s = \frac{1}{A_0}\int_{0}^{\infty}dx\,A_s(x,\omega) = \lambda_{\text{L}}
\,.
\end{equation}

For pure Nb we adopt the values of $\lambda_{\text{L$_0$}}\simeq 33\,\mbox{nm}$, $T_{c_0}=9.33\,\mbox{K}$ and $\Delta_0=1.55\,\mbox{meV}$ in Table~\ref{table-material_parameters}.~\cite{zar22}
Here we consider an N-doped Nb SRF cavities with a fundamental frequency, $f=\omega/2\pi=1.3\,\mbox{GHz}$, and transition temperature of $T_c=9.0\,\mbox{K}$.~\cite{baf21}
The combination of quasiparticle scattering by non-magnetic impurities and gap anisotropy leads to pair breaking, and accounts for the suppression of $T_c$ with a a quasiparticle scattering rate of $\tau\approx 2\times 10^{-13}\,\mbox{sec}$ [c.f. Fig.~12 of Ref.~\onlinecite{zar22}\,]. Thus, $\omega\tau\simeq 1.6\times 10^{-3}\ll 1$ as we assumed.
The normal-metal penetration depth with this level of disorder is then $\lambda_n\approx 
910\,\mbox{nm}$.
Disorder also suppresses the superfluid fraction; $n_s/n\approx 0.4$, and thus increases the zero-temperature London penetration depth to $\lambda_{\text{L}}\approx 52\,\mbox{nm}$.
For the $f=1.3\,\mbox{GHz}$ TM$_{010}$ mode of a cyclindrical cavity, the radius is $R=8.83\,\mbox{cm}$.
Thus, the difference in the penetration depth of the EM field for the normal and superconducting states implies a maximum {\it increase} in the cavity resonance frequency of order $\delta f\approx f\,(\lambda_n-\lambda_s)/R\approx 12.6\,\mbox{kHz}$, which is close to the maximum frequency shift for an N-doped Nb Tesla SRF cavity with $f=1.3\,\mbox{GHz}$, i.e. $\delta f_{\text{expt}}\simeq 12.5\,\mbox{kHz}$.~\cite{baf21} The geometric factor for an $f=1.3\,\mbox{GHz}$ a cylindrical cavity with $\delta f = 12.6\,\mbox{kHz}$ and radius $R=8.83\,\mbox{cm}$ corresponds to $G_{\text{cyl}}=268\,\Omega$, and thus $L\simeq 1.44\,R$, compared to $G_{\text{Tesla}}=270\,\Omega$.

While the maximum shift is positive, very near $T_c$ the cavity resonance frequency exhibits a negative frequency shift of order $\delta f \gtrsim -1\,\mbox{kHz}$ over a narrow temperature range near $T_c$ [c.f. Fig.4 of Ref.~\onlinecite{uek22}\,].
This anomaly can be qualitatively understood based on the divergence of the London penetration depth as $T\rightarrow T_c$, i.e. $\lambda_{\text{L}}(T)=\lambda_{\text{L}}/\sqrt{1-T/T_c}$. Thus, the temperature window over which the frequency shift is expected to be negative is $|\Delta T|/T_c<(\lambda_{\text{L}}/\lambda_{n})^2\approx 3.3 \times 10^{-3}$, which is in rough agreement with the experimentally observed width of the temperature anomaly of $|\Delta T|/T_c\vert_{\text{expt}}\simeq 3.8\times 10^{-3}$ at $f=1.3\,\mbox{GHz}$ [c.f. Fig.4 of Ref.~\onlinecite{uek22}\,].
More detailed and accurate analysis of the temperature dependence of the cavity resonance and quality factor, as well as an approximate analytical formula for the frequency anomaly very near $T_c$, follow.

\subsection{Impedance, Resistance and Reactance}

The EM response of cavity resonators, as well as 2D co-planar waveguide resonators, is often expressed in terms of the complex surface impedance, $Z(\omega)$. For SRF cavities the surface impedance resulting from the response of screening currents in the superconductor is defined by the ratio of the tangential fields at the interface,
\begin{equation}\label{eq-Zdef}
\hspace*{-3mm}
Z/Z_0
\equiv 
E_{||}(R)/H_{||}(R)
=
\frac{J_0(\varpi R/c)}{J_1(\varpi R/c)}
=
\frac{\varpi}{c}\,\Lambda(\varpi)^{-1}
\,.
\end{equation}
Thus, the solution to the boundary value problem, Eq.~\ref{eq-eigenvalue_equation}, directly determines the complex impedance. 
Equations~\eqref{eq-Z},~\eqref{eq-conductivity}, and~\eqref{eq-eigenvalue_equation} give $Z(\omega)$ directly in terms of the complex conductivity,
\begin{equation}\label{eq-Z}
Z = Z_0\sqrt{\frac{\omega}{4\pi i \sigma(\omega)}}
\,,
\end{equation}
and thus, $Z_s/Z_n=\sqrt{\sigma_n(\omega)/\sigma_s(\omega)}$.
This result is valid in the limit $c/\omega\gg\lambda_{s,n}$, which is well satisfied for SRF cavities at GHz frequecies.
The real and imaginary parts of $Z_{s/n}=R_{s/n}-i X_{s/n}$ are the surface resistance and surface reactance of the metal in the superconducting (s) and normal (n) states, respectively.
For SRF cavities in the normal state with a moderate level of disorder we have $\omega\tau\ll 1$, in which case $\sigma_{1n}\simeq\sigma_{\text{D}}$ and $\sigma_{2n}\simeq(\omega\tau)\sigma_{\text{D}}\ll\sigma_{1n}$. Thus, 
\begin{equation}\label{eq-normal-state_Rn}
X_n\simeq R_n\simeq Z_0\sqrt{\omega/8\pi\sigma_{\text{D}}}
\,.
\end{equation}

The change in the reactance upon cooling through the superconducting transition generates a shift in the resonance frequency, 
\begin{equation}
\label{eq-delta_f}
\delta f \equiv f_s-f_n 
= 
\frac{f}{2G}\,(X_n-X_s) 
\,,
\end{equation}
while the surface resistance below $T_c$ determines the quality factor,
\begin{equation}
\label{eq-Q}
Q_s=\frac{G}{R_s }
\,,
\end{equation}
where $G$ is the geometric factor of the cavity, which for the TM$_{010}$ mode of a cylindrical cavity is given by Eq.~\eqref{eq-G_cyl}.

\begin{figure}[!h]
\includegraphics[width=0.5\textwidth]{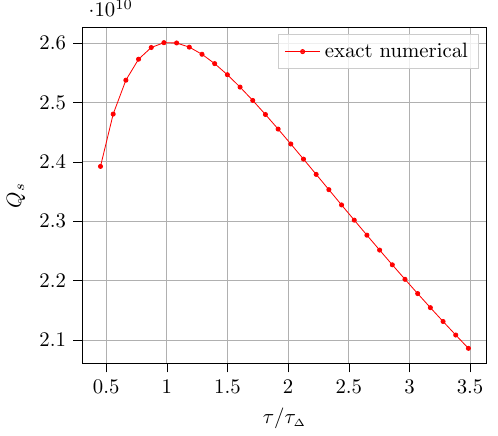}
\caption{Theoretical result for the quality factor $Q_s$ at $T/T_c=0.2$ 
as a function of the quasiparticle scattering time in units of $\tau_{\text{$\Delta$}}$, for a $f=2.6 GHz$ cavity.
}
\label{fig-Q_vs_tau}
\end{figure}

\subsection{Quality factor}

The effects of disorder on the quality factor $Q$ is important for development of SRF cavities for both acclerator and quantum applications.
Both the quality factor and resonance frequency of the cavity can be calculated directly from Eqs.~\eqref{eq-eigenvalue_equation} and \eqref{eq-K}.
In particular, for a cylindrical cavity of radius $R$ Eq.~\eqref{eq-eigenvalue_equation} reduces to
\begin{eqnarray}
\frac{1}{Q}
&=&
\frac{2}{R}\,\Im\left\{\frac{1}{\sqrt{K(\omega)}}\right\}
\\
\delta\omega
&=&
-\frac{\omega}{R}\,\Re\left\{\frac{1}{\sqrt{K(\omega)}}\right\}
\,,
\end{eqnarray}
in the limit $\delta\omega\ll\omega$ and $\omega/Q\ll\omega$, where the complex eigenvalue defined by the solution of Eq.~\eqref{eq-eigenvalue_equation} is written as $\varpi\equiv\omega+\delta\omega-i\omega/2Q$ with $\omega=x_{01}c/R$ being the TM$_{010}$ mode frequency of the ideal cavity. 

Figure~\ref{fig-Q_vs_tau} is shows the dependence of $Q$ on 
the quasiparticle scattering time in units of 
$\tau_{\text{$\Delta$}}\equiv\hbar/2\pi\Delta(T)$ 
for an $f=2.6 GHz$ Nb SRF cavity at $T=0.2T_c$. Thus, in the local limit, i.e. sufficient disorder such that $\lambda_{\text{L}}\gg\xi$, the maximum Q occurs for intermediate disorder with $\tau\sim\tau_{\text{$\Delta$}}$.
Indeed the non-monotonic dependence of $Q$ on $\tau$ is due to the pair-breaking suppression of the superfluid fraction, $n_s \propto 1/\lambda^2$, by quasiparticle-impurity scattering. 
For strong disorder, $\tau \ll \tau_{\text{$\Delta$}}$, the quality factor drops rapidly and approaches the $Q$ of the normal state for $\tau\rightarrow 0$.  
In the clean limit $\tau\rightarrow\infty$ the local approximation for the current response function breaks down for Nb. 
This is not an issue for current state of the art SRF cavities for accelerator applications, but it may be relevent to consider ultra-purity cavity technology for quantum sensors and processors in which case non-local electrodynamics of SRF cavities will be relevant. 

\begin{figure}[t]
\includegraphics[width=0.5\textwidth]{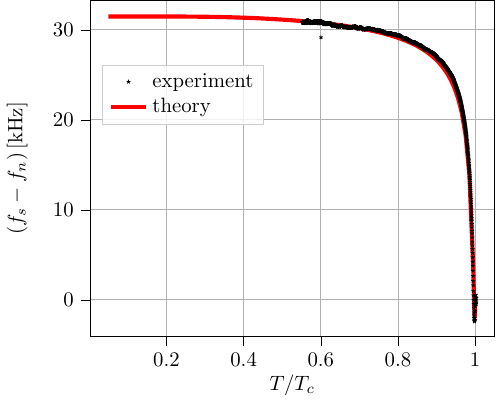}
\caption{
The frequency shift $\delta f=f_s-f_n$ as a function of temperature for a cavity with fundamental frequency $f=2.6 GHz$. The data (black symbols) are from Ref.~/onlinecite{baf21}. The theoretical result (red line), based on Eqs.~\eqref{eq-eigenvalue_equation} and~\eqref{eq-K}, corresponds to a quasiparticle scattering rate of $\tau =0.257\,ps$. Note the negative shift close to $T_c$.
}
\label{fig-delta_f_vs_T}
\end{figure}

\section{Comparison of Theory and Experiment}\label{sec-theory_vs_experiment}

Here we compare our results and analysis with measurements reported for N-doped Nb SRF cavities in Ref.~\onlinecite{baf21}. These authors also provide measurements of $T_c$ and the normal-state surface resistance, $R_n$, for Tesla cavities with mode frequencies of $f=\{0.65,1.3,2,6,3.9\}\,\mbox{GHz}$. 
When comparing theory with experimental data from Tesla cavities we rescale the magnitudes of $\delta f$ and $Q$ by replacing $G$ for the cylindrical cavity with the numerically calculated value of $G$ for the Tesla cavity.~\cite{baf21,uek22}

Figure~\eqref{fig-delta_f_vs_T} shows the temperature dependence of the frequency shift of a $f=2.6\,\mbox{GHz}$ cavity relative to the resonant frequency in the normal state just above $T_c$. The black symbols are the experimental data from Ref.~\onlinecite{baf21}. The calculated frequency shift is the solid red line. 
The theoretical curve corresponds to a quasiparticle scattering time of $\tau=2.57\times 10^{-13}\,\mbox{sec}$ ($\tau= 1.963\,\tau_0$).
This result is in reasonable agreement with the value of $\tau$ obtained from the reported normal-state suface resistance for the same cavity.
We estimate $\tau$ from $R_n$ using Eq.~\eqref{eq-normal-state_Rn} and the relations between the Drude conductivity, the plasma frequency and zero-temperature London penetration depth in the clean limit (c.f. Table~\ref{table-material_parameters}) to give,
\begin{equation}\label{eq-tau_vs_Rn}
\tau=\pi\,f\,\left(\frac{\lambda_{\text{L$_0$}}}{c}\,\frac{Z_0}{R_n}\right)^2
\,.
\end{equation}
Bafia et al.~\cite{baf21} report $R_n\simeq 7.1\,\times 10^{-3}\Omega$, which gives $\tau=0.28\,ps$, compared to the value of $\tau = 0.26\,ps$ obtained from our fit to the temperature-dependent frequency shift.
It is worth noting that there is no single value of the quasiparticle-impurity scattering rate, $1/\tau$, because the disorder is inhomogeneously distributed in the cavity walls. This fact is reflected in the distribution of values of $T_c$ measured at different locations on the cavity.~\cite{baf21}
We have analyzed the effects of inhomogeneity of $1/\tau$, and thus $T_c$, on the frequency shift and quality factor in Ref.~\onlinecite{uek22}.
The value of $1/\tau$ we obtain here from our fit to $\delta f(T,\tau)$ is in excellent agreement with the mean of the probability distribution $\rho(1/\tau)$ for the $2.6\,\mbox{GHz}$ N-doped Nb cavity.~\cite{uek22}
Results for other N-doped Nb cavities are summarized in Table~\ref{table-cavity_parameters} and discussed in detail in Ref.~\onlinecite{uek22}.

\begin{table}[t]
\begin{center}
\label{table-cavity_parameters}
\begin{tabular}{|c|c|c|c|c|c|}
\hline
$f[GHz]$ & 
$\tau[ps]$ & 
$\ell[nm]$ & 
$\tau/\tau_0$ 
& $R_n[m\Omega]$ & 
$\tau_{\text{R$_n$}} [ps]$
\\
\hline
0.65 & 0.173 & 45.092 & 1.327 & 4.37 & 0.184
\\
\hline                                        
1.30 & 0.224 & 58.235 & 1.713 & 5.45 & 0.236
\\
\hline                                        
2.60 & 0.257 & 66.700 & 1.963 & 7.10 & 0.279
\\
\hline                                        
3.90 & 0.249 & 64.730 & 1.905 & 8.96 & 0.262
\\
\hline
\end{tabular}
\caption{Material parameters for different cavities. The second column is the scattering time that gives the best fit of theory to experiment. The values of $R_n$ were provided by D. Bafia [private communication]. 
}
\end{center}
\end{table}

\subsection{Anomalous frequency shift}\label{sec-anomalous-frequency_shift}

\begin{figure}[t!]
\includegraphics[width=0.5\textwidth]{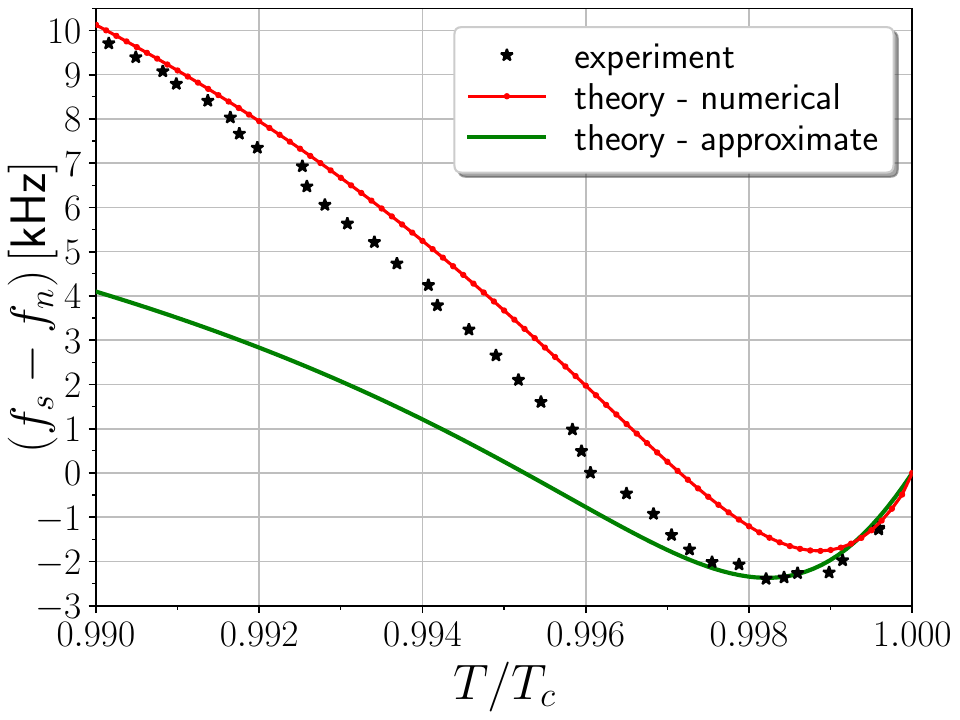}
\caption{
The experimental data for the frequency shift of the $2.6\,\mbox{GHz}$ cavity near $T_c$ are the black symbols. The minumum frequency shift is $-2.39\,\mbox{kHz}$ at $T=0.9982\,T_c$, and extends over a temperature window of $|\Delta T|\approx 4\times 10^{-3}\,T_c$. The theoretical result based on the best fit to the full temperature dependence of the frequency shift is the red line.
The green line is the approximate theoretical result based on the leading terms in the current response function as a function of $t=1-T/T_c$ with the constraint $\hbar\omega\ll 2\Delta(T)$ for the same scattering time, $\tau=0.257\,ps$.
\label{fig-df_vs_T-anomaly}
}
\end{figure}

Figure~\ref{fig-delta_f_vs_T} shows the negative frequency shift very close to $T_c$, which is expanded in Fig.~\ref{fig-df_vs_T-anomaly}. The region of negative frequency shift is where the London penetration depth, $\lambda_{\text{L}}(T)\approx \lambda_{\text{L}}/\sqrt{1-T/T_c}$ exceeds the normal-state penetration depth, $\lambda_n$. 
The experimental data is reasonably well described by theory given the fact that the scattering time $\tau$ was fit to the full temperature-dependent shift.

Here we provide an approximate analytic expression for the response function and the frequency shift anomaly near $T_c$.
%
%
From Eq.~\eqref{eq-conductivity} note that the real part of the current response function in Eq.~\eqref{eq-K} is proportional to the out of phase component of conductivity, and thus the penetration depth in the superconducting state,
\begin{equation}
\Re K(\omega) = \frac{4\pi\omega}{c^2}\,\sigma_2(\omega)
\xrightarrow[T\rightarrow T_c^-]{\hbar\omega\ll 2\Delta(T)}
\frac{1}{\lambda_{\text{L}}(T,\tau)^2}
\,,
\end{equation}
where 
$\lambda_{\text{L}}(T,\tau)\simeq\lambda_{\text{L}}/\sqrt{1-T/T_c}$ for 
$T\lesssim T_c$ for $T\simeq T_c^-$.
The imaginary part of $K(\omega)$ is proportional to the dissipative component of the conductivity,
\begin{equation}
\Im K(\omega) = -\frac{4\pi\omega}{c^2}\,\sigma_1(\omega)
\xrightarrow[T\rightarrow T_c^-]{\hbar\omega\ll 2\Delta(T)}
-\frac{1}{2\lambda_{\text{n}}(\omega,\tau)^2}
\,,
\end{equation}
with $\lambda_n(\omega,\tau)$ given by Eqs.~\eqref{eq-lambda_n} and \eqref{eq-delta}, i.e. we drop corrections of order $(\Delta(T)/\pi T_c)^2$ to the normal-state Drude conductivity.

Thus, for temperatures just below the superconducting transition, but still below the pair-breaking continuum, $\hbar\omega\ll 2\Delta(T)$, we obtain an approximate analytic form for the frequency shift and quality factor using Eq.~\eqref{eq-eigenvalue_equation} in the limit $\delta\omega/\omega \ll 1$ and $1/Q\ll 1$. In particular, the frequency shift of the superconductor \emph{relative to the ideal cavity frequency} for $|T-T_c|\ll T_c$ becomes,
\begin{equation}\label{eq-df_vs_T_approx}
\hspace*{-3mm}
\delta f_s = -\frac{f}{R_{\text{eff}}}\,
\Re
\left\{
\left(
\frac{1}
     {\lambda_{\text{L}}(T,\tau)^2}
     -\frac{i}{2\lambda_n(\omega,\tau)^2}
\right)^{-\frac{1}{2}}
\right\}
\,,
\end{equation}
where $R_{\text{eff}}=R\,G_{\text{Tesla}}/G_{\text{cyl}}$ and $R$ is the radius of a cylindrical SRF cavity with a TM$_{010}$ mode frequency of $f=2.6\,\mbox{GHz}$. The corresponding shift for the normal metallic state just above $T_c$ is $\delta f_n = -(f/R_{\text{eff}})\,\lambda_n$. Thus, the observable frequency shift of the superconducting cavity relative to the normal state is $\delta f = \delta f_s - \delta f_n$.
This approximate result reproduces the negative frequency shift near $T_c$, including an good estimate of the temperature, and minimum value, of the negative frequency shift, but deviates from the full theory as the temperature drops further. Thus, the origin of the negative shift is the competition between the large normal-state penetration depth for disordered Nb and the London penetration depth which competes with $\lambda_n$ only as $T\rightarrow T_c^-$.

\section{Conclusion}

Theoretical results for the frequency shift and quality factor, as well as the suppression of $T_c$ and superfluid density by disorder, can provide powerful analysis tools for characterizing disorder in SRF cavities for both accelerator and detector applications.

\section{Acknowledgements}

We thank Daniel Bafia, Anna Grassellino, Alex Romanenko and John Zasadzinski for discussions on their results on N-doped Nb SRF cavities, and for motivating this study.
This work was supported by the U.S. Department of Energy, Office of Science, National Quantum Information Science Research Centers, Superconducting Quantum Materials and Systems Center (SQMS) under contract number DE-AC02-07CH11359.

%
\end{document}